\documentclass[%
 reprint,
 amsmath,amssymb,
 aps,
prl,
]{revtex4-1}

\usepackage{graphicx}
\usepackage{dcolumn}
\usepackage{bm}
\usepackage{enumerate}   
\usepackage{xcolor}
\usepackage{hyperref}

\usepackage{lineno}
\begin{document}

\preprint{APS/123-QED}

\title{New constraints for QCD matter from improved Bayesian parameter estimation in heavy-ion collisions at LHC}

\author{~J.E.~Parkkila$^{1,2}$}%
\author{~A.~Onnerstad$^{1,2}$}
\author{~S.~F.~Taghavi$^{3}$} 
\author{~C.~Mordasini$^{3}$} 
\author{~A.~Bilandzic$^{3}$} 
\author{~D.J.~Kim$^{1,2}$}%
\email{jasper.parkkila@cern.ch}

\affiliation{$^{1}$University of Jyv\"askyl\"a, Department of Physics, P.O. Box 35, FI-40014 University of Jyv\"askyl\"a, Finland}
\affiliation{$^{2}$Helsinki Institute of Physics, P.O.Box 64, FI-00014 University of
Helsinki, Finland}
\affiliation{$^{3}$Physik Department, Technische Universit\"{a}t M\"{u}nchen, Munich, Germany}



\date{\today}

\begin{abstract}
The transport properties of quark-gluon plasma created in relativistic heavy-ion collisions are quantified by an improved global Bayesian analysis using the CERN Large Hadron Collider Pb--Pb data at $\sqrt{s_{\textbf{NN}}}=2.76\;$ and $5.02\;$TeV. The results show that the uncertainty of the extracted transport coefficients is significantly reduced by including new sophisticated collective flow observables from two collision energies for the first time. This work reveals the stronger temperature dependence of specific shear viscosity, a lower value of specific bulk viscosity, and a higher hadronization switching temperature than in the previous studies. The sensitivity analysis confirms that the precision measurements of higher-order harmonic flow and their correlations are crucial in extracting accurate values of the transport properties.
\end{abstract}

\maketitle


\renewcommand{\vec}[1]{\mathbf{#1}}

\newcommand{\trento}{T\raisebox{-.5ex}{R}ENTo}

\section{\label{sec:PRLsingle_section}}

\raggedbottom 
The experiments utilizing ultra-relativistic heavy-ion collisions (HIC) play an important role in understanding many-body Quantum Chromodynamics (QCD). The high center-of-mass energy of heavy-ion collisions at the Relativistic Heavy Ion Collider (RHIC) and Large Hadron Collider (LHC) liberates the confined quarks and gluons inside nuclei to form a medium called quark-gluon plasma (QGP) \cite{Borsanyi:2013bia,HotQCD:2014kol,Braun-Munzinger:2015hba,Busza:2018rrf}. In the past years, phenomenological multi-stage models (containing initial, pre-equilibrium, QGP, hadron gas stages) have given a solid description of heavy-ion physics. In particular, the QGP stage is successfully explained by causal relativistic hydrodynamics with two first-order transport coefficients, namely the shear and bulk viscosity over entropy density ($\eta/s$ and $\zeta/s$, respectively). The comparison of model predictions with measurements indicates that the experimental data favor small values for $\eta/s$ and $\zeta/s$, which implies that the produced QGP in HIC is considered the most perfect fluid observed in nature~\cite{Bernhard2019}. 
The formed QGP is in the strongly coupled regime, in which the applications of the perturbative techniques are limited. On the other hand, the non-perturbative techniques (i.e. gauge/gravity duality and lattice QCD) are restricted to specific scenarios \cite{Kovtun:2004de,Bazavov:2019lgz,Meyer:2007ic,Astrakhantsev:2017nrs,Meyer:2007dy,Astrakhantsev:2018oue}. Consequently, accurate experimental measurements to constrain these quantities are crucial to deepen our understanding of QCD.

To this date, the number of free parameters (including temperature-dependent $\eta/s(T)$ and $\zeta/s(T)$) in a typical multi-stage heavy-ion collision model ranges from 10 to 20. Considering only few of these parameters can be estimated theoretically, they must be extracted from the experimental observations, e.g., particle yields, anisotropy in final particle distribution in momentum space, particle mean transverse momentum, etc.~\cite{Aamodt:2010cz, Abelev:2013vea, ALICE:2011ab}. The free parameters usually have a complex relationship with the experimental observables, such that inferring the parameter values from the experimental data is not an easy task. In this respect, a substantial progress has happened in recent years by employing Bayesian analysis. In addition to the seminal works in Refs.~\cite{Bernhard:2015hxa,Bernhard:2016bar,Bernhard:2016tnd,Bernhard:2018hnz,Bernhard2019} on applying the Bayesian analysis in heavy-ion physics, other studies have been done in which few extra experimental observables are employed to infer the parameters and/or few variations of multi-stage models are considered~\cite{Auvinen:2020mpc,Nijs:2020ors,Nijs:2020roc,JETSCAPE:2020mzn}.

Among the possible experimental observables, some of them are more sensitive to the properties of the system controlling the details of its collective evolution. For instance,
it has been demonstrated that symmetric cumulants (see Ref.~\cite{Bilandzic:2013kga}) are sensitive to $\eta/s(T)$~\cite{ALICE:2016kpq,ALICE:2017kwu}. These quantities belong to a larger class of experimental observables used to quantify the anisotropic flow, which is one of the most informative experimental probes in heavy-ion physics (see also Refs.~\cite{Borghini:2000sa,Borghini:2001vi,ALICE:2010suc,Jia:2014jca,DiFrancesco:2016srj,Mordasini:2019hut,Bilandzic:2020csw,Taghavi:2020gcy,Bilandzic:2021rgb}). In this letter, we start with the same multi-stage model as in Ref.~\cite{Bernhard2019}, but in contrast to the observables used in that work, we employ the new observables that were measured only recently by ALICE experiment in Pb--Pb collisions at two collision energies to increase our sensitivity to hydrodynamic transport coefficients $\eta/s$  and $\zeta/s$. To this end, we include symmetric cumulants~\cite{ALICE:2016kpq,ALICE:2017kwu,ALICE:2021adw}, generalized symmetric cumulants~\cite{ALICE:2021klf}, and flow harmonic mode couplings~\cite{Acharya:2020taj} as the input in our Bayesian analysis. The experimental measurements for particle yields and particle mean transverse momentum at $\sqrt{s_{\textbf{NN}}}=5.02\;$TeV ~\cite{Acharya:2019yoi,Adam:2016ddh} are added to increase our sensitivity on the collision energy dependence of the model. We employ identical methods in extracting the observables of interest from the output of simulations to the ones which were used in the corresponding experimental measurements, in order to avoid any incompatibilities in comparison. As our main result, we report an improved estimation for $\eta/s(T)$ and $\zeta/s(T)$  as well as the improved sensitivity of the anisotropic flow estimations to the model parameters.

\textit{Model parameters, experimental observables and Bayesian analysis approach.---}In the present study, the model setup is mainly identical with Refs.~\cite{Bernhard2019,Parkkila:2021tqq}. The {\tt \trento{}} model~\cite{Moreland:2014oya} is used for the initial conditions. At the pre-equilibrium stage, free streaming connects the initial state to the QGP stage. The system evolution continues in this deconfined stage via a 2+1 causal hydrodynamic model, {\tt VISH2+1}~\cite{Shen:2014vra,Song:2007ux}. The temperature dependence of the shear and bulk viscosities over entropy density are parameterized as the following: 
\begin{equation}
	\label{eq:etaparam}
	(\eta/s)(T) = (\eta/s)(T_c)+(\eta/s)_\mathrm{slope}(T-T_c) \left(\frac{T}{T_c}\right)^{(\eta/s)_\mathrm{curve}},
\end{equation}
and
\begin{equation}
	\label{eq:zetaparam}
	(\zeta/s)(T) = \frac{(\zeta/s)_\mathrm{max}}{1 + \left(\frac{T-(\zeta/s)_{T_\mathrm{peak}}}{(\zeta/s)_\mathrm{width}}\right)^2}.
\end{equation}
A particlization model switches the partonic degrees of freedom to hadrons \cite{Pratt:2010jt,Bernhard:2018hnz}. The evolution in the hadron gas continues with the UrQMD model~\cite{Bass:1998ca,Bleicher:1999xi}. We have tabulated 14 different parameters of these models in Table~\ref{tab:design} with their corresponding prior range, the optimal MAP-value (Maximum A Posteriori), as well as a short description. The only difference of our setup compared to Ref.~\cite{Bernhard2019} is that one common centrality definition is shared between all prior parametrizations, unlike in Ref.~\cite{Bernhard2019}, where the centrality was defined individually for each parametrization by sorting the resulting events into centrality bins. However, our initial condition prior range is narrow, and we do not expect to see large multiplicity variations that would cause bias due to shared centrality definition. Furthermore, for each event, we sample the hypersurface exactly ten times regardless of the cumulative number of particles. 

\begin{table*}[tbh!]
  \caption{
    \label{tab:design}
    Input parameter ranges for the initial condition and hydrodynamic models. 
  }
  \begin{tabular}{p{2.2cm}  p{7.1cm} p{3.0cm} p{1.2cm} }
    \hline\hline
    Parameter         & Description                        & Range    & MAP       \\
    \hline
    N(2.76 TeV)              & Overall normalization (2.76 TeV)              & [11.152,\,18.960] & 14.373 \\
    N(5.02 TeV)              & Overall normalization (5.02 TeV)              & [16.542,\,25] & 21.044 \\
    $p$               & Entropy deposition parameter       & [0.0042 ,\,0.0098]    & 0.0056 \\
    $\sigma_k$ & Std. dev. of nucleon multiplicity fluctuations & [0.5518,\,1.2852] & 1.0468 \\
    $d_{\min}^3$ & Minimum volume per nucleon & [$0.889^3$,\,$1.524^3$] & $1.2367^3$\\
    $\tau_\mathrm{fs}$ & Free-streaming time & [0.03,\,1.5] & 0.71\\
    $T_c$ & Temperature of const. $\eta/s(T)$, $T < T_c$ & [0.135,\,0.165] & 0.141\\
    $\eta/s(T_c)$  & Minimum $\eta/s(T)$  & [0,\,0.2]  & 0.093     \\
    $(\eta/s)_\mathrm{slope}$ & Slope of $\eta/s(T)$ above $T_c$ & [0,\,4] & 0.8024\\
    $(\eta/s)_\mathrm{curve}$ & Curvature of $\eta/s(T)$ above $T_c$ & [$-1.3$,\,$1$] & $0.1568$\\
    $(\zeta/s)_\mathrm{peak}$  & Temperature of $\zeta/s(T)$ maximum & [0.15,\,0.2] & 0.1889\\
    $(\zeta/s)_{\max}$ & Maximum $\zeta/s(T)$ & [0,\,0.1] & 0.01844\\
    $(\zeta/s)_\mathrm{width}$ & Width of $\zeta/s(T)$ peak & [0,\,0.1] & 0.04252\\
    $T_\mathrm{switch}$ & Switching / particlization temperature & [0.135,\,0.165] & 0.1595\\
    \hline\hline
  \end{tabular}
\end{table*}

The Bayesian analysis is a powerful tool to obtain the model parameters from the experimental measurements. In the following, we briefly explain its main steps and refer the reader to Ref.~\cite{Bernhard:2018hnz} for more details. We represent a generic set of the model parameters and output observables by vectors $\vec{x}$ and $\vec{y}$, respectively. Considering we have poor knowledge about the free parameters initially, our degree of belief on the parameter values is encoded into a uniform \textit{prior} distribution $P(\vec{x})$ in intervals defined in Table~\ref{tab:design}. According to the Bayes' theorem, the updated degree of belief in the light of experimental data (\textit{posterior} distribution) is given by $P(\vec{x}|\vec{y})\propto P(\vec{y}|\vec{x}) P(\vec{x}) $. The probability $P(\vec{y}|\vec{x})$, the likelihood, is obtained by probing the parameter space $\vec{x}$ and comparing it with experimental measurements $\vec{y}$. Markov Chain Monte Carlo (MCMC) method is employed to probe the parameter phase space to obtain the posterior distribution via Bayes's theorem.  Given that heavy-ion models are computationally expensive, instead of using the model directly, the computations are done on 500 parameter design points distributed with Latin hypercube scheme~\cite{TangHypercube,MORRIS1995381}. At each designed point, $3\times 10^6$ events are generated for the 5.02~TeV collision energy, and $5\times 10^6$ for the 2.76 TeV, including the ten samples of the hypersurface. The Gaussian process (GP) is used to emulate the model in a continuous parameter phase space. The predictions in between the design points have been validated.

The following measurements from ALICE experiment have been used in Ref.~\cite{Bernhard2019}: centrality dependence of charged and identified particles yields $d\mathrm{N}/\mathrm{d}y$, mean transverse momentum $\langle p_\mathrm{T} \rangle$~\cite{ALICE:2010mlf,ALICE:2013mez,ALICE:2015juo,ALICE:2016igk,ALICE:2014gvd}, as well as two-particle anisotropic flow coefficients $v_n\{2\}$ for harmonics $n=2$, 3, and 4~\cite{ALICE:2011ab,ALICE:2016ccg}. In the present study, besides the recent measurements for identified particle yields and $\langle p_\mathrm{T} \rangle$ at $\sqrt{s_{\textbf{NN}}}=5.02\;$TeV \cite{Adam:2016ddh,Acharya:2019yoi} that have not been used in the previous study, we employ latest measurements related to the anisotropic flow: two-particle anisotropic flow coefficients $v_n\{2\}$ for $n=5,\ldots,9$ \cite{ALICE:2016ccg,Acharya:2017gsw,Acharya:2020taj}, normalized symmetric cumulants NSC$(k,\ell)$~\cite{ALICE:2016kpq,ALICE:2017kwu,ALICE:2021adw}, and flow mode couplings $\chi_{n,mk}$ \cite{ALICE:2017fcd,Acharya:2020taj}. In a previous study in Ref.~\cite{Parkkila:2021tqq}, only measurements at $\sqrt{s_{\textbf{NN}}}=5.02\;$TeV has been considered, while measurements from both collision energies $\sqrt{s_{\textbf{NN}}}=2.76\;$TeV and 5.02\;TeV is implemented into this analysis. In particular, the latest measurements of the generalized normalized symmetric cumulants NSC$(k,\ell,m)$ at $\sqrt{s_{\textbf{NN}}}=2.76\;$TeV~\cite{ALICE:2021klf} are included.

The methods used for the calculations of the observables are the same as the experimental analysis in Refs.~\cite{ALICE:2016kpq,Acharya:2017gsw,Acharya:2020taj,ALICE:2021adw}. 
In order to obtain internally consistent comparison, the centrality classes for this study were chosen in such a way that they match the centrality classes of the experimental data.
The multiplicity range has to be defined for each centrality class. This is done by using the MAP parametrization from~\cite{Bernhard2019} to simulate events and select the resulting minimum bias events by charged-particle multiplicity $\mathrm{d}N/\mathrm{d}\eta$ at midrapidity ($|\eta| < 0.5$). 
By counting and averaging the particle species at midrapidity, we could evaluate the identified particle multiplicity $\mathrm{d}N/\mathrm{d}\eta$ and $\langle p_\mathrm{T}\rangle$.
For the experimental data there is no additional processing required for the preparation of the comparison, since it is already corrected and extrapolated to zero $p_\mathrm{T}$~\cite{Abelev:2013vea}.
Our model only reproduces the spectra of protons for the identified $\mathrm{d}N/\mathrm{d}\eta$, hence they were the only species used for the model calibration.
With this information we can calculate the flow coefficients and other observables for charged particles within the acceptance of the ALICE detector, using the same methods as in~\cite{Acharya:2020taj, ALICE:2021adw}.

\color{black}

\begin{figure}[tbh!]
	\centering
	\includegraphics[width=1.0\linewidth]{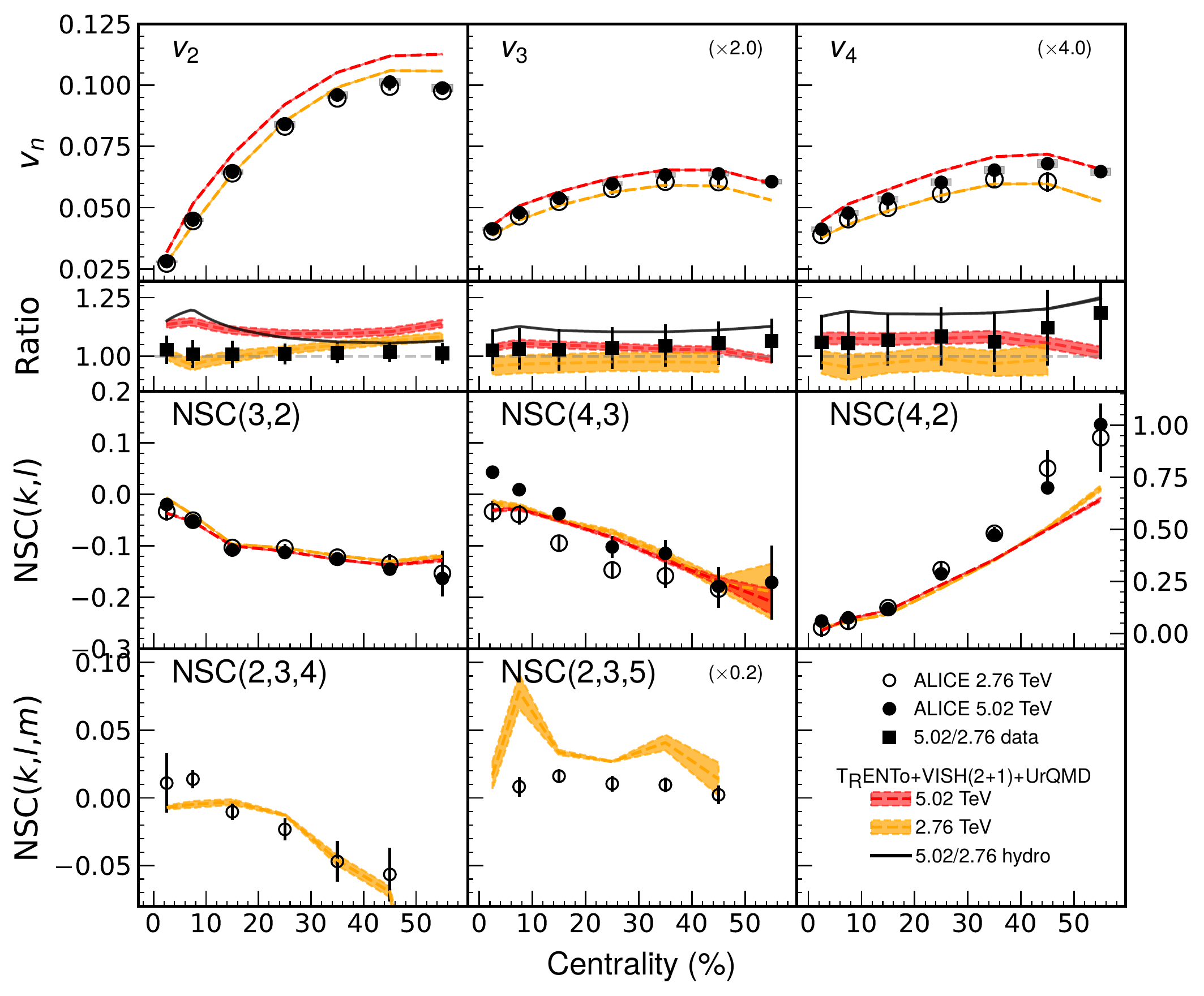}
	\caption{(color online) Flow coefficients $v_n$ and normalized symmetric cumulants ($\mathrm{NSC}(k,l)$ and $\mathrm{NSC}(k,l,m)$) from two hydrodynamical calculations using the MAP parametrization are compared to the experimental data~\cite{Acharya:2020taj,ALICE:2021adw}.
	The red band corresponds to the calculations at the collision energy of $\sqrt{s_{\text{NN}}}=5.02\,\mathrm{TeV}$, while the yellow band presents the results at $\sqrt{s_{\text{NN}}}=2.76\,\mathrm{TeV}$. The corresponding ratio between the data and calculation for the respective collision energies is shown for the $v_n$. Here, the black markers and black lines are the ratio between the two collision energy results, for data and calculations, respectively.}
	\label{fig:nsc_result}
\end{figure}

As it is mentioned before, a uniform prior distribution is considered for the parameters. Since the new observables included in this study should be more sensitive to the transport coefficients, we assume that the parameters of the initial state model are uniformly distributed around the MAP values found in Ref.~\cite{Bernhard2019}. A narrow range of variations is allowed for further minor adjustments.

\begin{figure}[tbh!]
	\centering
	\includegraphics[width=1.0\linewidth]{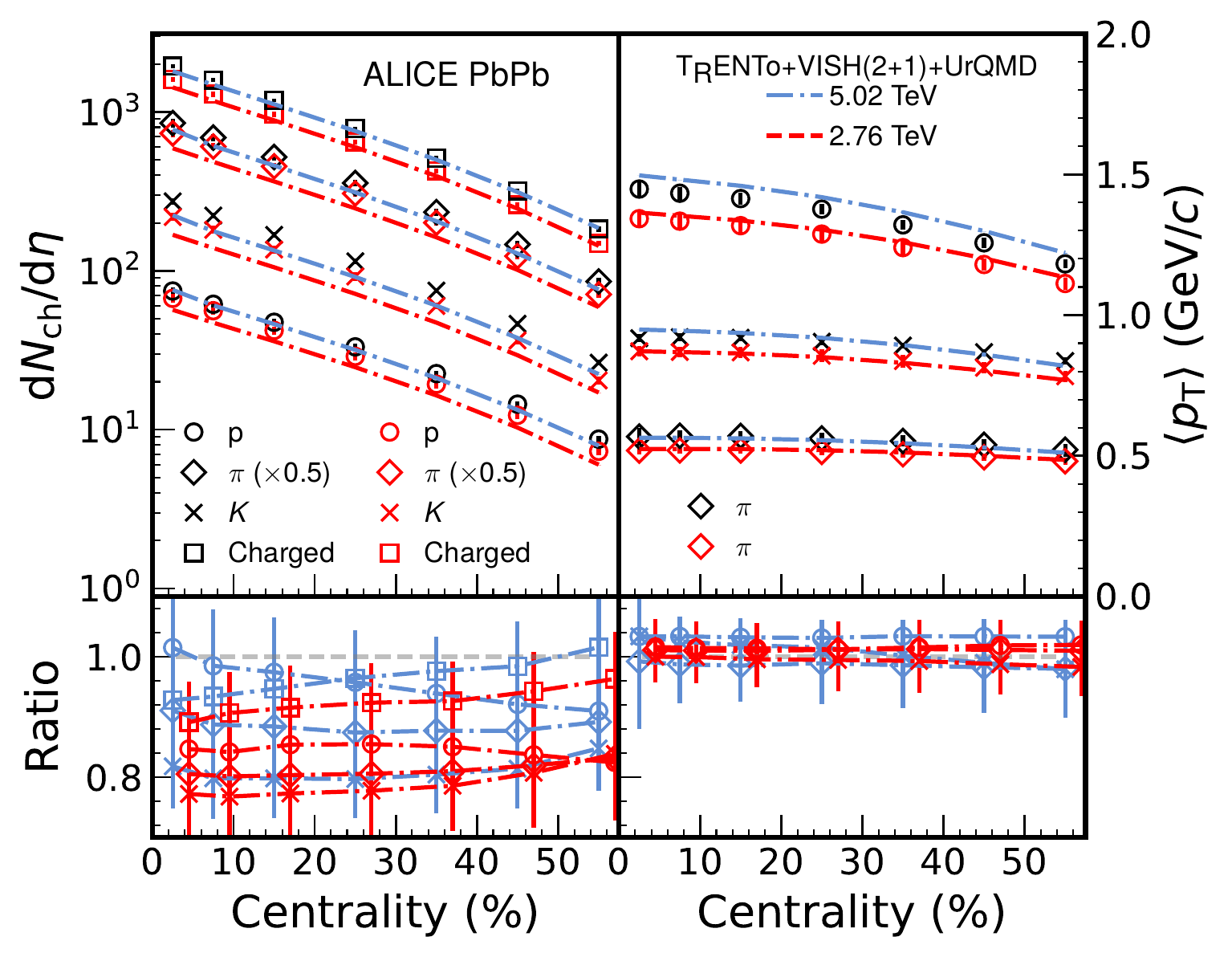}
	\caption{(color online) Charged and identified particle multiplicity and $\langle p_\mathrm{T} \rangle$ from two hydrodynamical calculations are compared to the experimental data at center-of-mass energy of 2.76 and 5.02 TeV.}
	\label{fig:mult_meanpt_MAP}
\end{figure}

\begin{figure}[tbh!]
	\centering
	\includegraphics[width=1.0\linewidth]{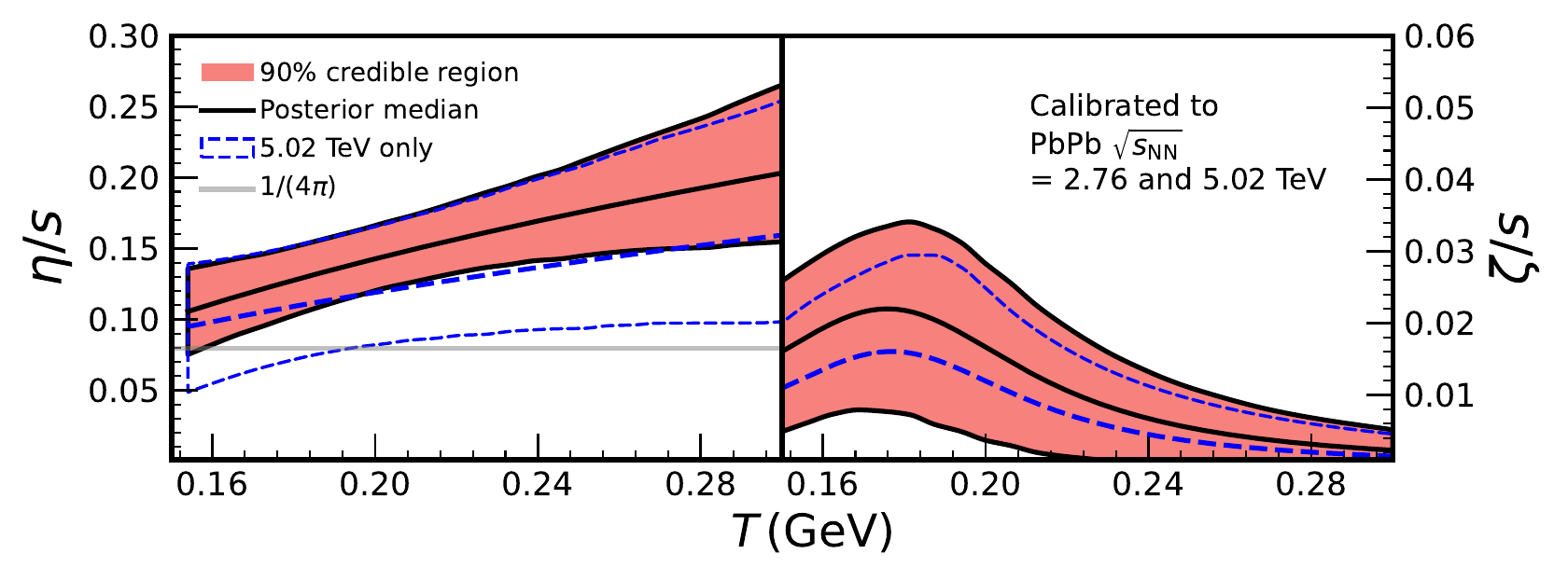}
	\caption{The 90\%-credibility region for the shear (top) and bulk (bottom) viscosity to entropy ratio is given as a red band. The black line represents the median of the credibility range. Our result is compared to the MAP parametrization from~\cite{Parkkila:2021tqq}, for which the calibration was performed using 5.02 TeV data only.}  
	\label{fig:regions}
\end{figure}

\begin{figure*}[tbh!]
	\centering
	\includegraphics[width=1.0\linewidth]{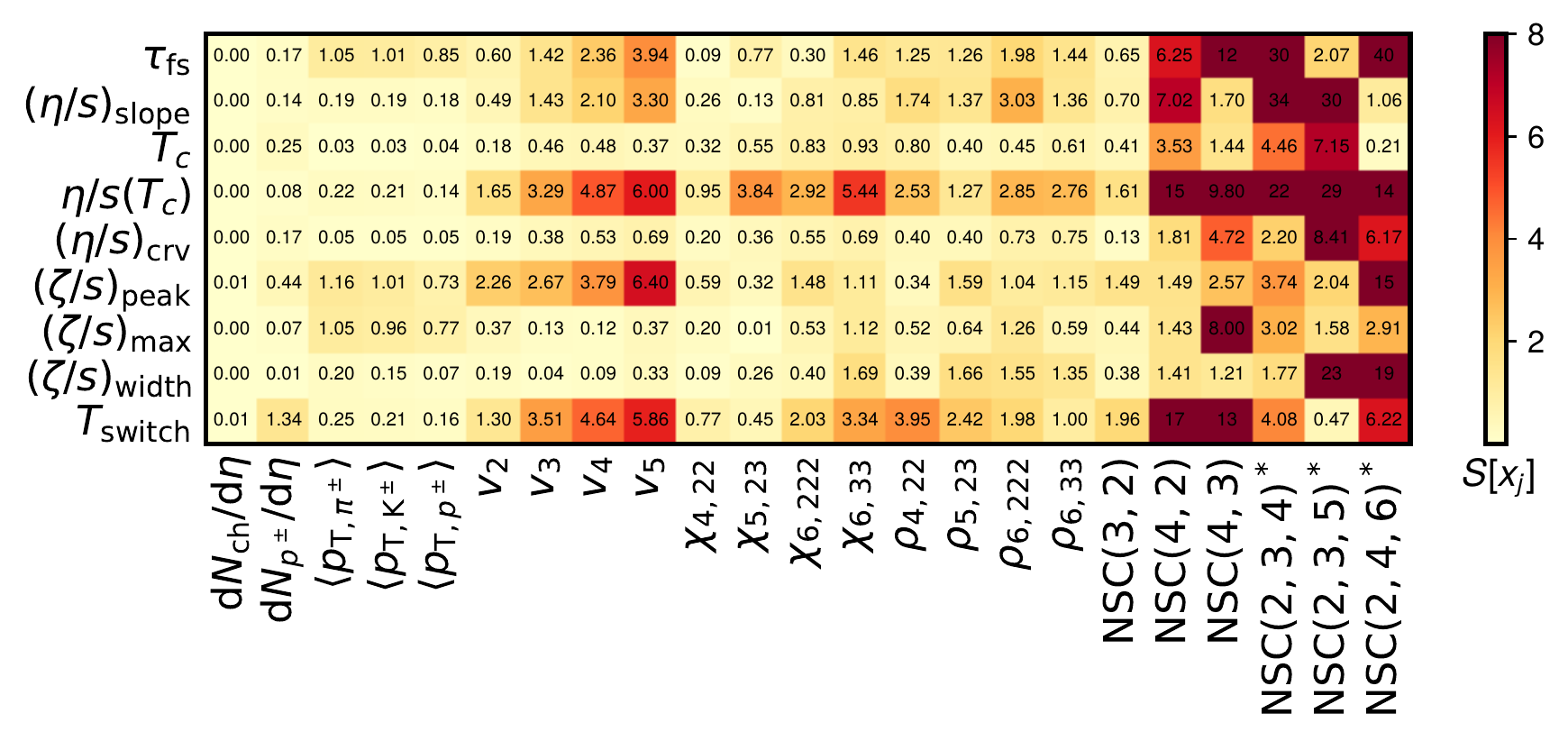}
	\caption{Sensitivity of the observables to the model parameters visualized as a color map. The asterisk ($^*$) for an observable indicates that the sensitivity was evaluated using 2.76 TeV calculations, whereas the rest are evaluated using 5.02 TeV calculations. The sensitivity index is averaged over four centrality classes, from 5\% to 40\%, except for NSC($k$,$l$,$m$), for which only one centrality class 20-30\% is used. Light yellow shades represent a very limited sensitivity or no sensitivity, whereas orange and darker red colors represent moderate or strong sensitivities to the corresponding model parameter, respectively.}
	\label{fig:sensitivity_obs}
\end{figure*}

\textit{Results and discussion.---}After finding the posterior distribution $P(\vec{x}|\vec{y})$, we extract those values of  $\vec{x}$ that maximize the distribution (MAP values). In Fig.~\ref{fig:nsc_result}, the model predictions for observables related to the anisotropic flow are compared with the measurements. As seen from the figure, the overall trend of the data is captured by the model. The observables indicate a different dependence on the collision energy in the simulation than experimental measurements. The difference between two energies is clearly visible in the centrality dependence of $v_2$, where the predictions for most central collisions are significantly larger than for peripheral collisions. The experimental measurements for $v_2\{2\}(5.02~\text{TeV})/v_2\{2\}(2.76~\text{TeV})$ (black filled markers in the ratio panel) is compatible with unity in a wide range of centrality classes, while the simulation (black curve in the same panel) reaches 25\% above unity in some centralities. The ALICE measurement reveals a sign change for NSC$(4,3)$ at $\sqrt{s_{\text{NN}}}=5.02\,\mathrm{TeV}$ in central collisions, while there is no sign change in $\sqrt{s_{\text{NN}}}=2.76\;$TeV measurement. We do not observe such a collision energy-dependent behavior in the simulation. One notes that the only collision energy-dependent part of the model is considered to be the overall initial energy density normalization. The simulation also fails to explain data at peripheral collisions for NSC$(4,2)$. All results considered, the higher energy description is found to be worse for all observables, except for $v_5$, $\chi_{6,222}$, and proton, pion and charged particle multiplicity based on the same $\chi^2$-test performed in~\cite{Parkkila:2021tqq}.

Switching temperature, $T_{\text{switch}}$, is the temperature at which the hydrodynamic evolution of QGP changes from the deconfined stage into the hadron-gas stage. Including the new observables raises the previous estimation for $T_{\text{switch}}$ from  $\approx$~$0.150\,\mathrm{MeV}$ reported in Ref.~\cite{Bernhard:2018hnz} to $\approx$~$0.160\,\mathrm{MeV}$. It has been discussed in Refs.~\cite{Acharya:2017gsw,ALICE:2017fcd,Acharya:2020taj} that the newly added anisotropic flow observables, mode couplings and correlation between harmonics are sensitive to the viscous corrections to the equilibrium distribution at the freeze-out~\cite{Luzum:2010ad,Luzum:2010ae,Teaney:2012ke,Yan:2015jma}.

The centrality dependence of charged and identified particle yields and $\langle p_\mathrm{T} \rangle$ is shown in Fig.~\ref{fig:mult_meanpt_MAP}. The model predictions with MAP parametrization are shown by red and blue curves for the center-of-mass energies of 2.76~TeV and 5.02~TeV, respectively. As seen from the figure, the simulation does not lead to an accurate prediction for charged and identified particle yields for both energies. For particle yields, the predictions and measurements are in better agreement at the center-of-mass energy 5.02~TeV. Together with what has been observed for $v_2\{2\}$ measurements at central collisions, these discrepancies can be considered as evidence that we need a revision on our understanding about the model collision energy dependence.

In Fig.~\ref{fig:regions}, the temperature dependence of $\eta/s$ and $\zeta/s$ are presented. The result for $\eta/s(T)$ agrees with that reported in Ref.~\cite{Bernhard2019}. Compared to the previous analysis with $\sqrt{s_{\text{NN}}}=$5.02~TeV data only \cite{Parkkila:2021tqq}, an improvement in  the uncertainty of $\eta/s(T)$  is observed. Moreover, this parameter shows a stronger temperature dependence than in the previous study, meaning we observe a more substantial departure from the lower bound $1/4\pi$. We also find higher mean values for $\zeta/s(T)$. Including both 2.76~TeV and 5.02~TeV center-of-mass energy data improves the uncertainty of $\zeta/s(T)$. As it is mentioned earlier, the symmetric cumulants are sensitive to the temperature dependence of $\eta/s$. Our new observation in $\zeta/s(T)$ uncertainty improvement indicates that the newly added anisotropic flow observables including normalized symmetric cumulants are sensitive to the temperature dependence of $\zeta/s$ as well. In the following, we study the parameter sensitivity more systematically.


To compare the sensitivity of the observables with each other, we follow Refs.~\cite{JETSCAPE:2020mzn,Hamby:1994} and define the sensitivity of an observable $\hat{O}$ to the parameter $x_j$ via $S[x_j]=|\hat{O}(\vec{x}')-\hat{O}(\vec{x})|/\delta \hat{O}(\vec{x})$ where $\hat{O}(\vec{x})$ is the value of the observable at the parameter point $\vec{x}=(x_1,\ldots,x_p)$. The quantity $\vec{x}'$ is a point in the parameter space with a small difference in a single parameter $x_j$, $\vec{x}'=(x_1,\ldots,(1+\delta)x_j,\ldots,x_p)$. The small quantity $\delta$ is chosen to be equal to 0.1. We have found that the larger values for $\delta$ lead to similar results. The result is depicted in Fig.~\ref{fig:sensitivity_obs}.  
As seen from the figure, compared to the other observables, the normalized symmetric cumulants NSC$(k,\ell)$ and the generalized normalized symmetric cumulants NSC$(k,\ell,m)$ are very sensitive to the values of transport coefficient parameters. This result is more general and more quantitative evidence of what has been observed in Refs.~\cite{ALICE:2016kpq,ALICE:2017kwu} for the sensitivity of SC$(k,\ell)$ to $\eta/s$. Here, we indicate that NSC observables are sensitive to both  $\eta/s$ and $\zeta/s$. An interesting feature that we immediately recognize from Fig.~\ref{fig:sensitivity_obs} is that by considering the higher harmonics and higher-order cumulants, the shear and bulk viscosity parameters modifications reveal more drastic change on the observables. For temperature-independent $\eta/s$, it has been shown that higher harmonics have more sensitivity to $\eta/s$ modification~\cite{Alver:2010dn,Teaney:2012ke}. This study has been generalized to temperature-dependent $\eta/s$ for $v_2$ and $v_3$ by Gardim and Ollitrault~\cite{Gardim:2020mmy}.
The effect can be understood as follows: the higher harmonics capture finer details of initial state energy density structures. The dissipation effects should wash out the finer structures during hydrodynamic evolution. As a result, small changes in the value of $\eta/s$ and $\zeta/s$ affect the higher harmonic observables more drastically. The high sensitivity of NSCs cannot be merely due to high harmonic flow coefficients, since the mode coupling observables contain the same harmonics but show less sensitivity. We deduce that the genuine correlations between flow amplitudes $v_n$, captured by NSCs, are particularly sensitive to the transport properties of the medium.


\textit{Summary and outlook.---}Building on the previous studies, we employed the latest measurements of higher harmonics, higher-order flow fluctuation observables as inputs into a Bayesian analysis. The present study indicated that these observables are sensitive to the transport coefficients
and revealed the importance of the precision measurements of these observables to infer the hydrodynamic transport coefficients accurately. Including the latest flow harmonic measurements, we have improved the uncertainty of estimated values for $\eta/s$ and $\zeta/s$. Despite using the new observables as inputs to extract model parameters, there are remaining discrepancies between model and experimental measurements. For instance, NSC(4,2) model prediction is improved in our new analysis, but it still deviates from measurements at higher centralities. At $\sqrt{s_{\text{NN}}}=5.02\,\mathrm{TeV}$, the sign change of NSC(4,3) in the lower centralities is not reproduced neither in Ref.~\cite{Bernhard2019} nor in our study. Further investigations are needed in this respect. These discrepancies, together with poor model/data compatibility for the energy scale dependence of $v_2\{2\}$ at central collisions and also the particle yields, show the necessity to improve our understanding of the heavy-ion collision models.

\acknowledgments
\textit{Acknowledgments.---} We thank Jonah E. Bernhard, J. Scott Moreland and Steffen A. Bass for the use of their viscous relativistic hydrodynamics software and their valuable comments on various processes of this work. We would like to thank Harri Niemi, Kari J. Eskola and Sami R\"as\"anen for fruitful discussions. We acknowledge Victor Gonzalez for his crosscheck for various technical parts of the event generation.
We acknowledge CSC - IT Center for Science in Espoo, Finland, for the allocation of the computational resources. 
This research was completed using $\sim 64$ million CPU hours provided by CSC. 
Three of us (SFT,CM, and AB) have received funding from the European Research Council (ERC) under
the European Unions Horizon 2020 research and innovation program (Grant Agreement No. 759257).
\nocite{*}

\bibliography{apssamp}

\newpage

\section{Supplemental Material}

\begin{figure}[tbh!]
\centering
\begin{minipage}{.45\textwidth}
	\centering
	\includegraphics[width=1\textwidth]{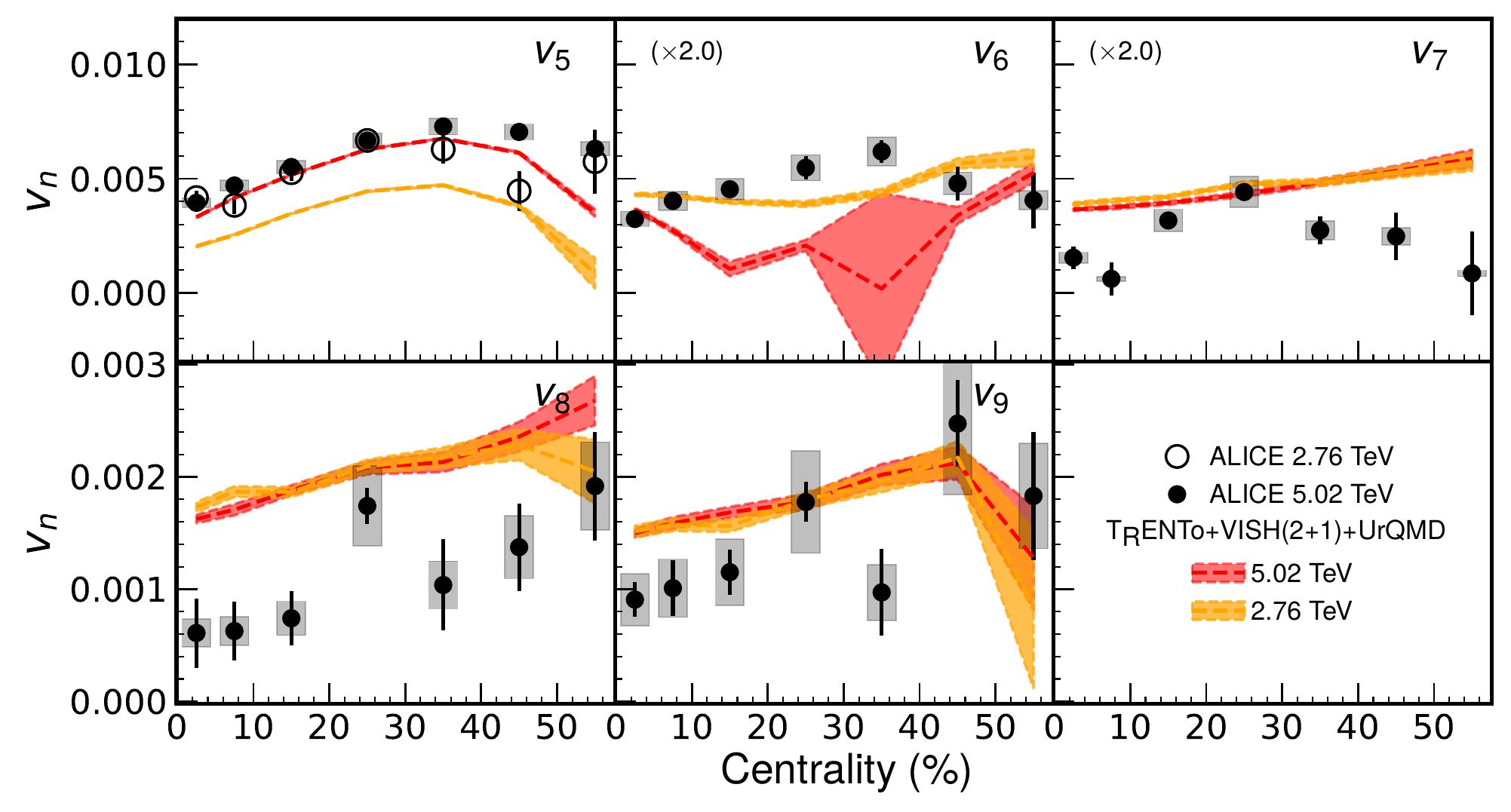}
	\caption{(color online) Flow coefficients for harmonics 5 to 9. The red and yellow bands present the model prediction for collisions energies $\sqrt{s_{\text{NN}}}=5.02$ and $2.76\;$TeV, respectively. The experimental data are published in Refs.~\cite{ALICE:2017fcd,Acharya:2020taj}. }
	\label{fig:v5Tov9}
\end{minipage}%
\hspace{0.02\textwidth}
\begin{minipage}{.45\textwidth}
	\centering
	\includegraphics[width=1\textwidth]{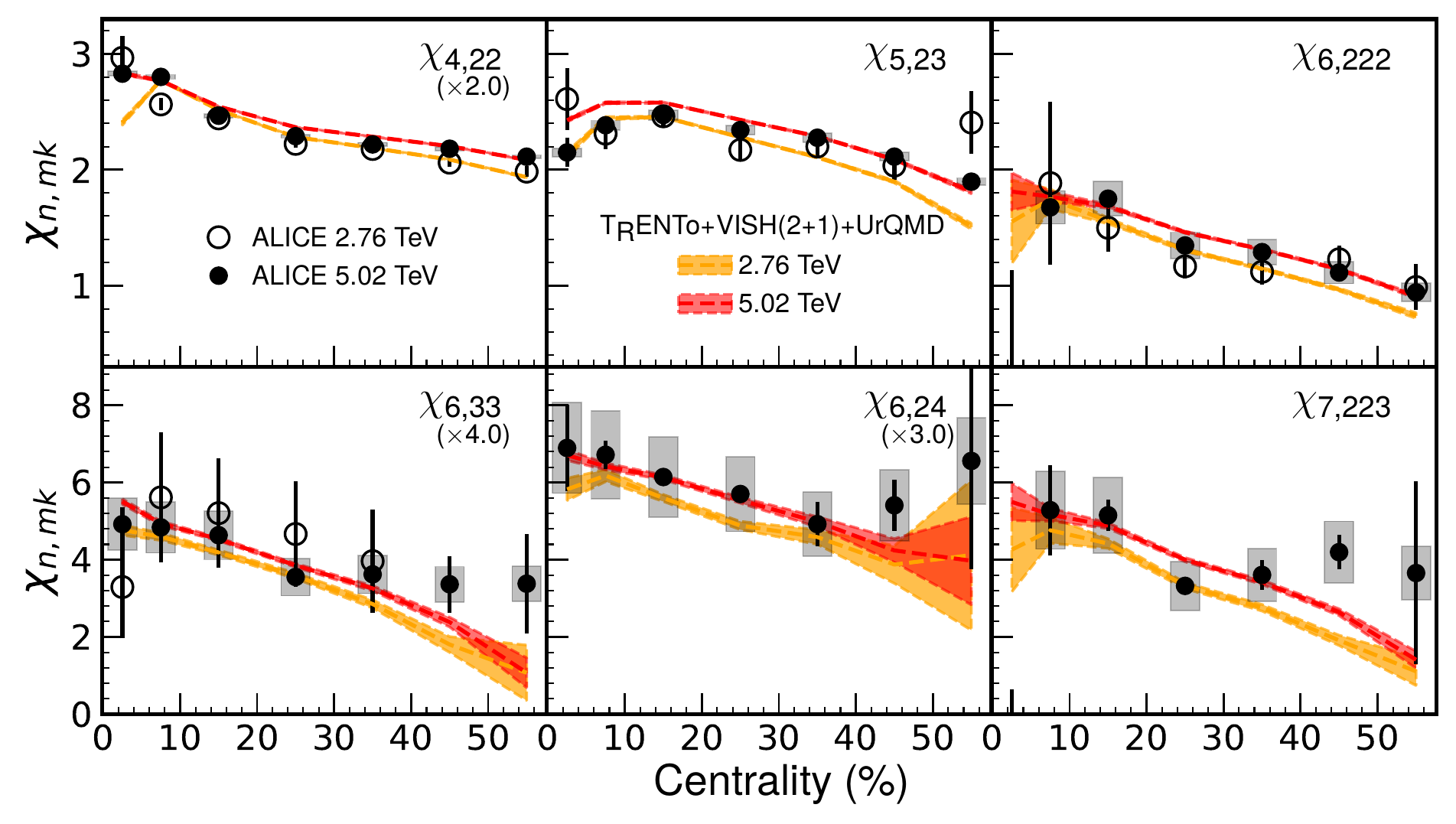}
	\caption{(color online) Flow mode couplings for six different harmonic combinations. The experimental data are published in Refs.~\cite{ALICE:2017fcd,Acharya:2020taj}. }
	\label{fig:chi}
\end{minipage}
\begin{minipage}{.45\textwidth}
	\centering
	\includegraphics[width=1\textwidth]{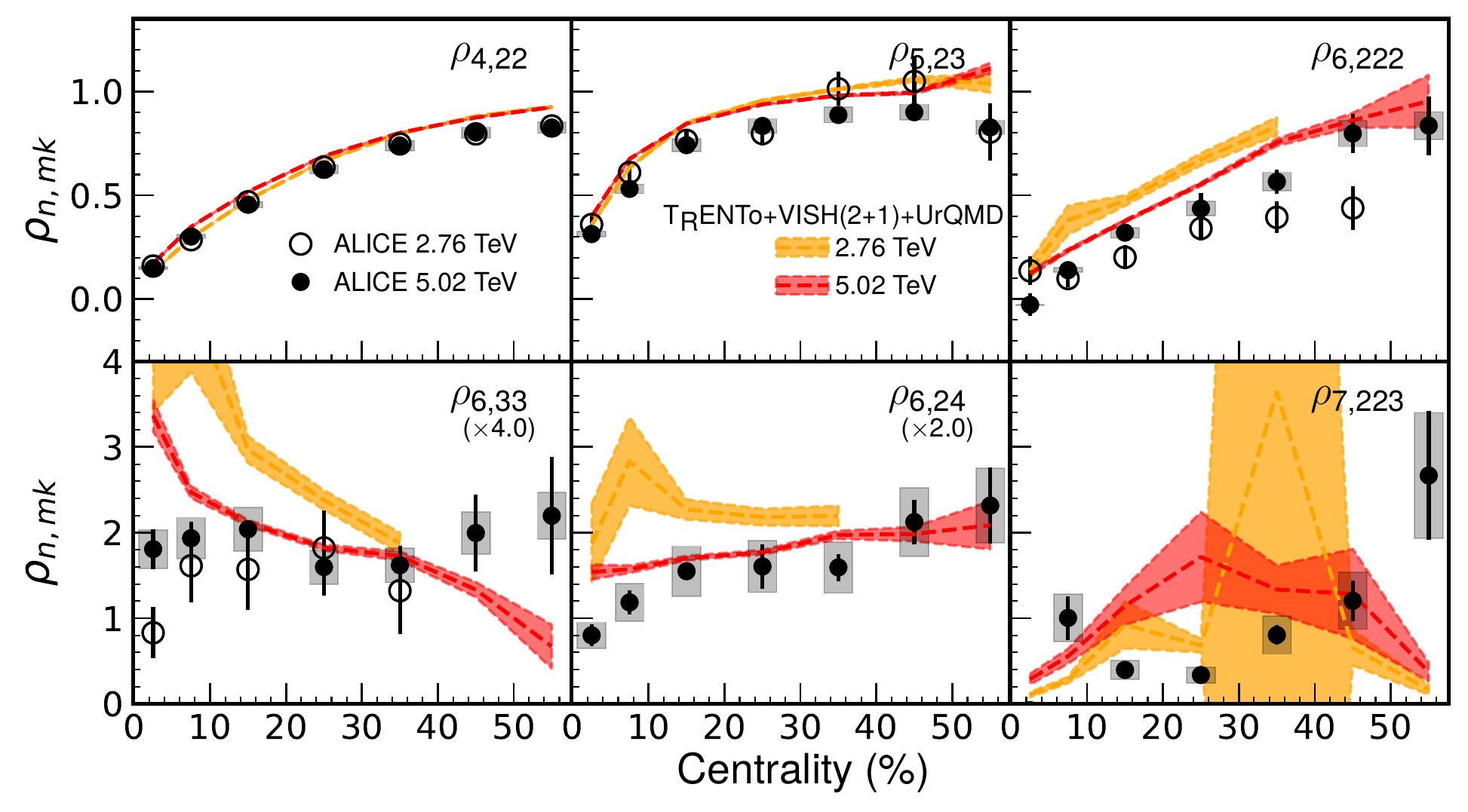}
	\caption{(color online) Symmetry plane correlations for six different harmonic combinations. The experimental data are published in Refs.~\cite{ALICE:2017fcd,Acharya:2020taj}. }
	\label{fig:rho}
\end{minipage}
\end{figure}

This supplemental material presents extra information about the model predictions with MAP parameterization and posterior distribution of the model parameters.

In the main paper, the model predictions for charged and identified particle yields, $\langle p_T \rangle$, and a few anisotropic flow observables have been compared with the measurements (see Fig.~\ref{fig:nsc_result} and Fig.~\ref{fig:mult_meanpt_MAP}). Here, we present the comparison between simulation and data for additional anisotropic flow observables. The flow cumulants $v_n\{2\}$ for $n=5,\ldots,9$, flow mode couplings and symmetry plane correlations for various harmonics are presented in Figs.~\ref{fig:v5Tov9}--\ref{fig:rho}, respectively. As seen from the figures, although the overall trends are compatible with the measurement, the model does not accurately explain data for harmonic six and above. We observe more compatibility between simulation and data in mode-coupling observables, even in cases that higher harmonic flow coefficients are involved.

\begin{figure}[tbh!]
	\centering
	\includegraphics[width=1\linewidth]{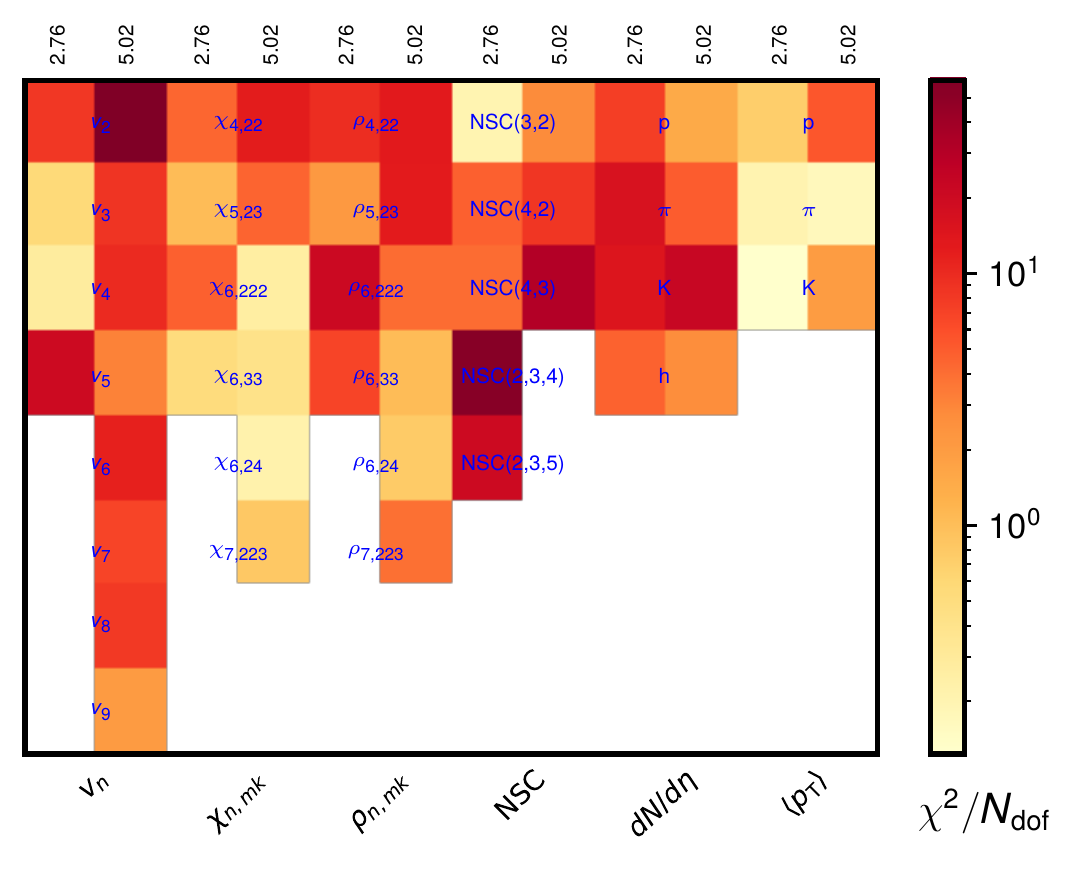}
	\caption{The $\chi^2$-test values calculated between the data and model calculations for both beam energies are shown for all flow harmonic mode couplings, symmetric cumulants, generalized symmetric cumulants, and, charged and identified particle multiplicity and $\langle p_\mathrm{T} \rangle$.}
	\label{fig:chi2all}
\end{figure}

Figure~\ref{fig:chi2all} presents the same $\chi^2$-test as in~\cite{Parkkila:2021tqq} to quantify the agreement of the models with the data for the 0--60\% centrality range. In addition to the flow harmonic mode couplings and symmetric cumulants, the generalized symmetric cumulants, particle multiplicity and $\langle p_\mathrm{T} \rangle$ were added to the test. These results show that the higher energy description are worse for all observables except for $v_5$, $\chi_{6,222}$, and charged particle multiplicities.

\begin{figure}[tbh!]
	\centering
	\includegraphics[width=0.9\linewidth]{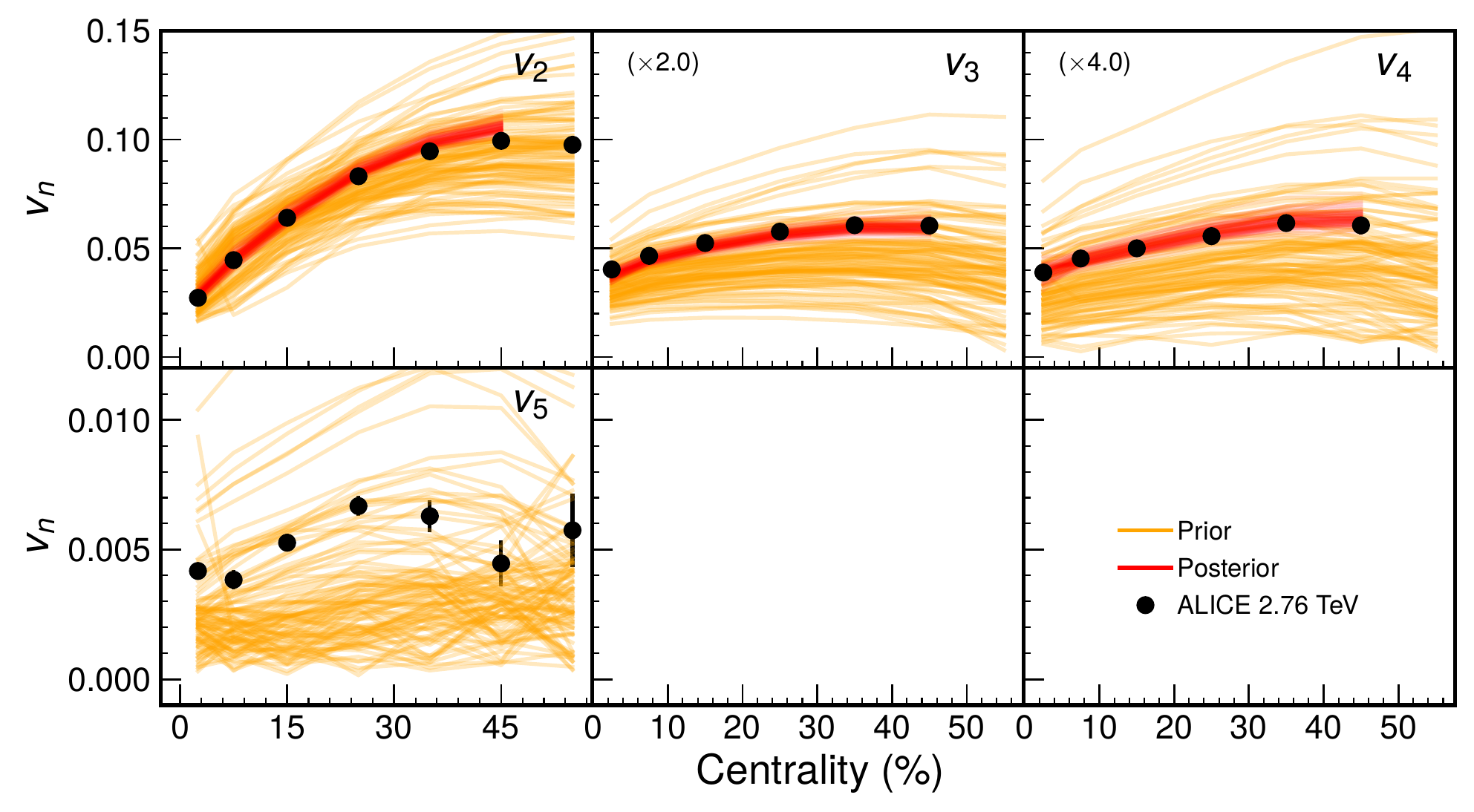}
	\caption{Flow coefficients $v_n$ as given by the design parametrizations are presented in yellow curves. All harmonics are simultaneously covered by the design parametrizations. The red curves represent a number of curves sampled from the posterior distribution, and as given by the emulator.}
	\label{fig:vn_params}
\end{figure}

\begin{figure}[tbh!]
	\centering
	\includegraphics[width=0.7\linewidth]{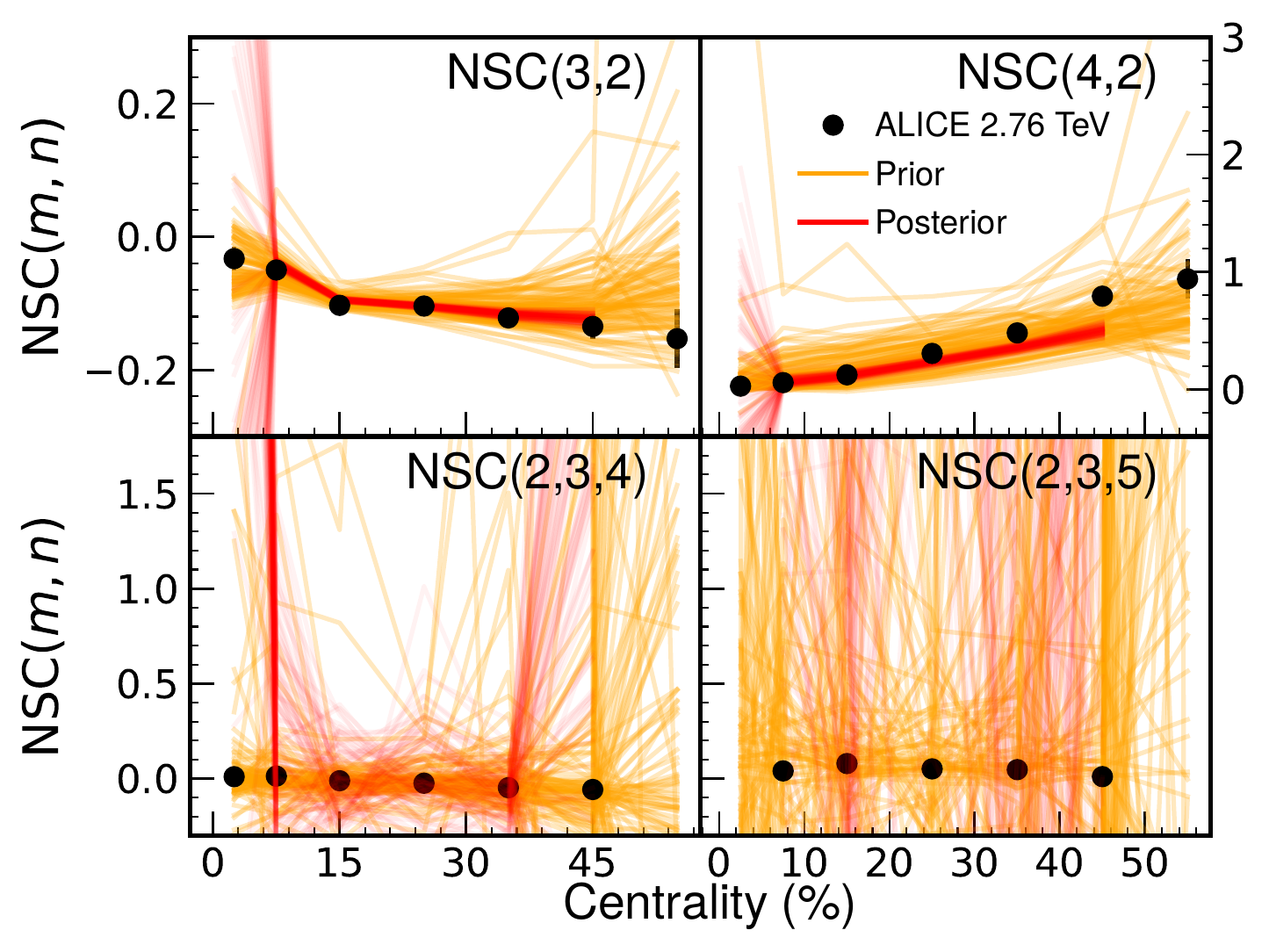}
	\caption{Design parametrizations for normalized symmetric cumulants (in yellow) and a number of posterior sample curves as given by the emulator (in red).}
	\label{fig:nsc_params}
\end{figure}

The model calculations using the design parametrizations obtained from the prior distribution for each observable at $\sqrt{s_{\textbf{NN}}}=2.76\;$TeV (see Ref.~\cite{Parkkila:2021tqq} for 5.02\;TeV) are shown in Figs.~\ref{fig:vn_params}--\ref{fig:nsc_params}. The yellow curves represent the calculations corresponding to each design parametrization point which are used in training the GP emulator. The red curves are from the GP emulator predictions corresponding to random points sampled from the posterior distribution. 

\begin{figure}[tbh!]
	\centering
	\includegraphics[width=0.7\linewidth]{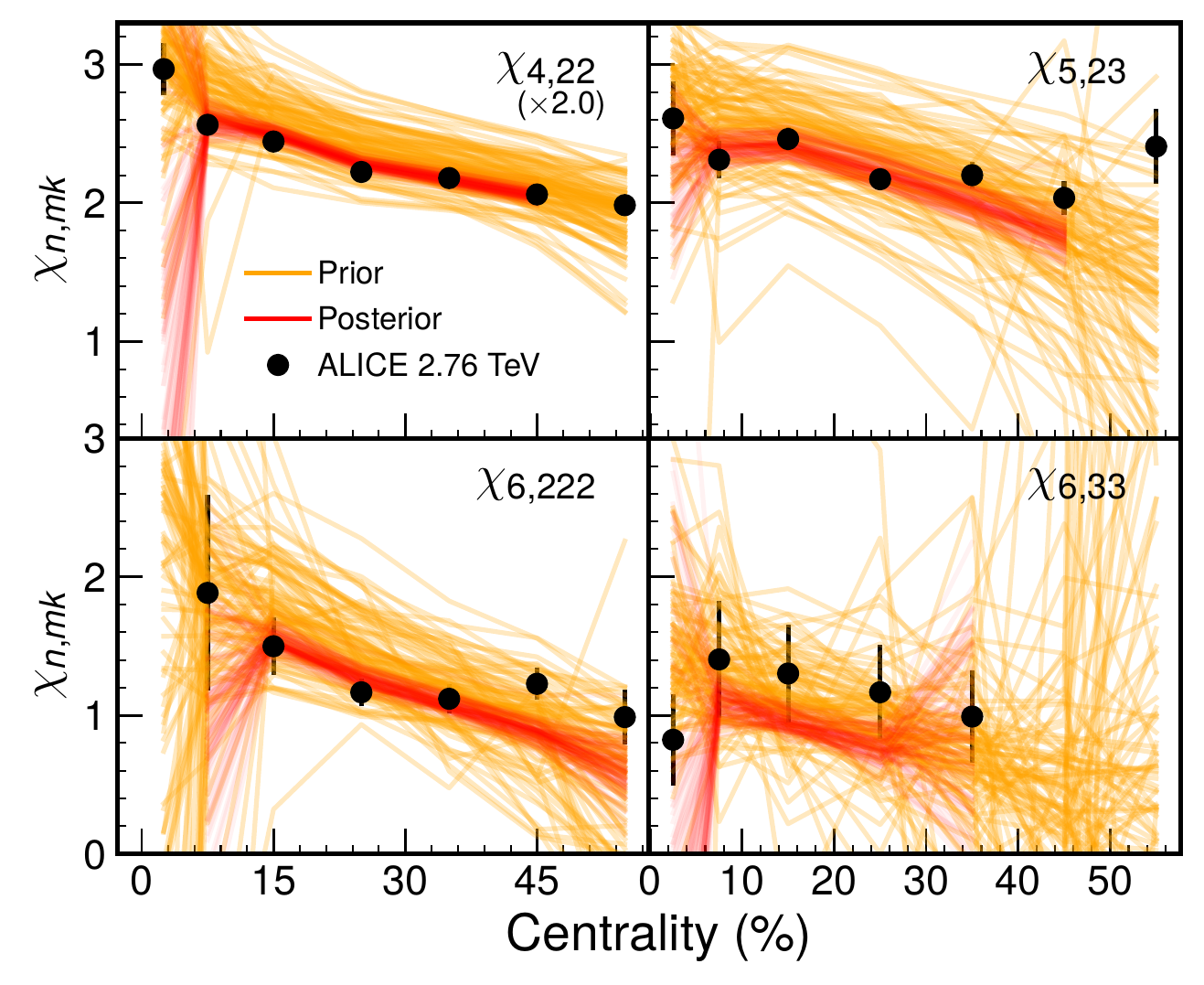}
	\caption{Design parametrizations for non-linear flow mode coefficients $\chi_{n,mk}$ (in yellow) and a number of posterior sample curves as given by the emulator (in red).}
	\label{fig:chi_params}
\end{figure}

\begin{figure}[tbh!]
	\centering
	\includegraphics[width=0.7\linewidth]{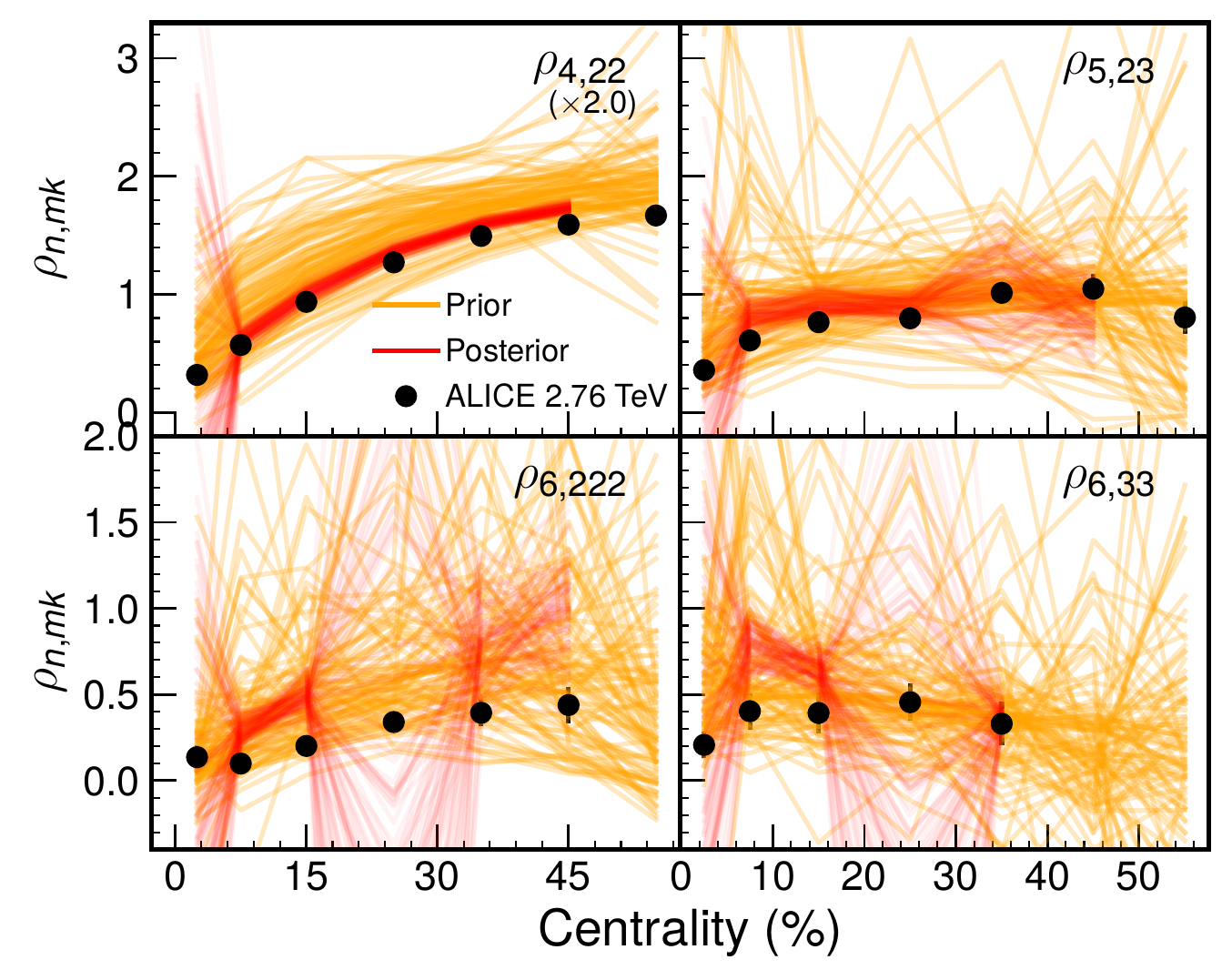}
	\caption{Design parametrizations for symmetry plane correlations $\rho_{n,mk}$ (in yellow) and a number of posterior sample curves as given by the emulator (in red).}
	\label{fig:chi_params}
\end{figure}

\begin{figure*}[tbh!]
	\centering
	\includegraphics[width=1\linewidth]{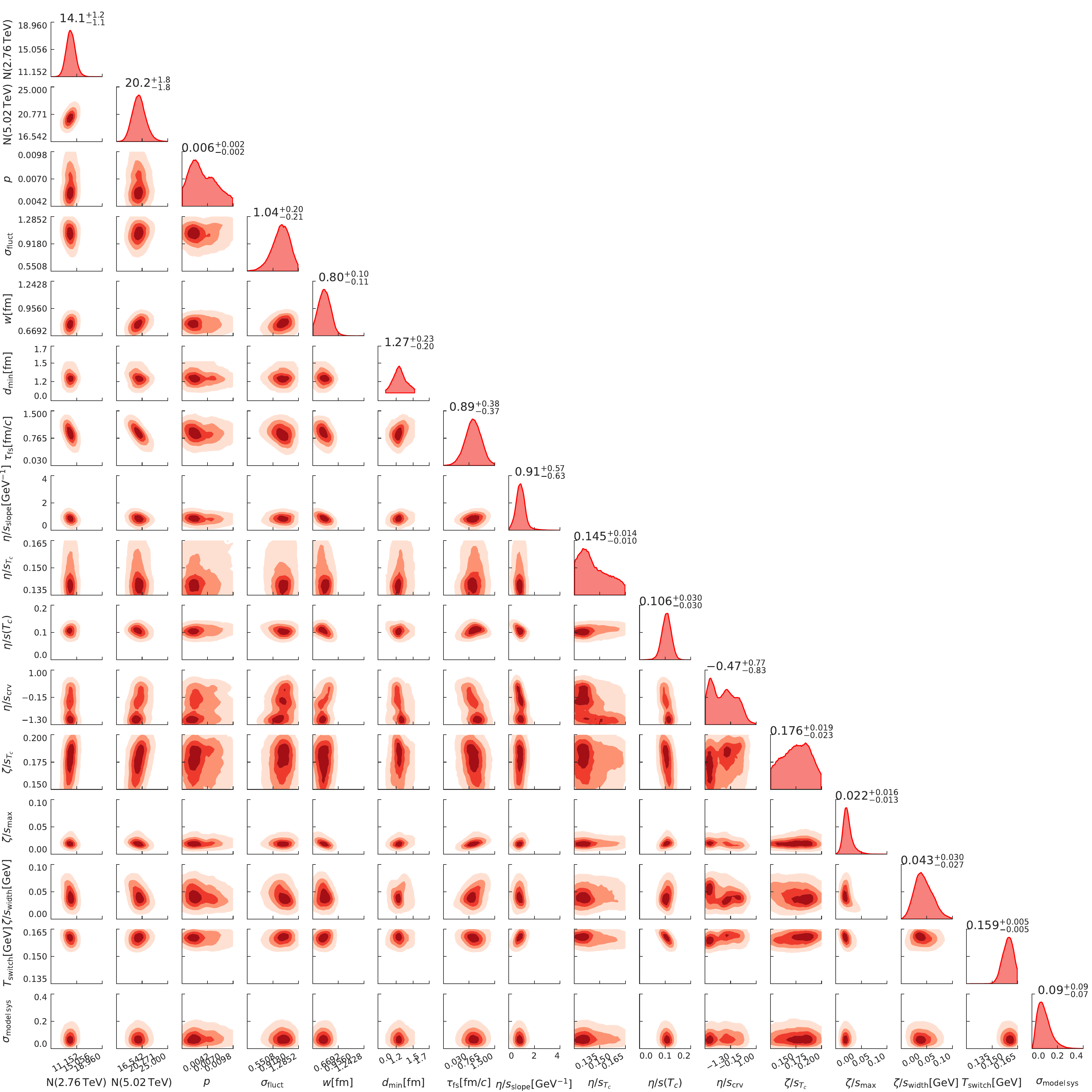}
	\caption{Marginal and joint marginal parts of the poster distribution. The numbers denoted on top of marginal distributions are the median together with the range of 90\%  credibility.  }
	\label{fig:marginal_distribution}
\end{figure*}

The MAP values for the model parameters are presented in Table~\ref{tab:design}, which are the median of the marginal posterior distribution for a given parameter. For the readers interested in more information about the posterior distribution, we present the marginal (diagonal panels) and joint marginal (off-diagonal panels) part of the posterior distribution in Fig.~\ref{fig:marginal_distribution}. The results are compatible with previous studies in Refs.~\cite{Bernhard2019,Parkkila:2021tqq}. However, focusing on parameters related to $\eta/s(T)$ and $\zeta/s(T)$, we find that the parameters are inferred with more accuracy as we expect. For instance, we can see a more sharp peak for parameter $(\zeta/s)_{\text{peak}}$. The marginal distribution of this parameter was more broadened in the previous studies. Moreover, the joint marginal distribution between parameters $(\zeta/s)_{\text{peak}}$ and $(\zeta/s)_{\text{curve}}$ is concentrated in a smaller region of the parameter space.

\end{document}